\documentclass[a4paper]{jpconf}
\usepackage{graphicx,amssymb}

\begin{document}
	
\title{Studying patched spacetimes for binary black holes}

\author{J Tarrant$^1$ and G Beck$^1$}

\address{$^1$ School of Physics, University of the Witwatersrand, Johannesburg, WITS-2050, South Africa}

\ead{justine.tarrant@wits.ac.za}

\begin{abstract}
Circumbinary accretion disks have been examined, theoretically, for supermassive and intermediate mass black holes, however, disks for black hole masses in the LIGO regime are poorly understood. Assuming these binaries possess such a disk initially, the question we want to answer is: are they dissipated by outflows or accretion prior to inspiral? To study this problem we propose a novel approach, whereby we consider an approximate, analytic spacetime and solve the geodesic equation for particles in this spacetime so that we can determine the likely fate of particles coming from the accretion disk. Preliminary indications suggest a likelihood of accretion prior to inspiral.
\end{abstract}

\section{Introduction}
An interesting astrophysical problem to consider is the treatment of circumbinary accretion disks (hereafter, disks) around binary black holes. Currently, in the literature, supermassive ($10^5 - 10^9 M_{\odot}$) and intermediate ($10^2 - 10^5 M_{\odot}$) mass black holes have been considered with disks. The case of disks around binaries in the LIGO (Laser Interferometer Gravitational wave Observatory) regime (tens of solar masses) are more obscure. It is worth noting that circumbinary disks have not been observed for any mass regime. Examining these problems comes with numerical challenges since one generally employs hydrodynamical and magneto-hydrodynamical simulations which are computationally expensive and may be impractical \cite{Mundim_2014}. To date, numerical relativity remains one of the most powerful probes used to study the strong regime in general relativity around extreme environments such as binary black holes and their mergers \cite{baumgarte}. 

Numerical simulations must proceed from initial data which is tricky to produce \cite{cook}. One way to calculate the initial data is to construct an approximate, analytic, global metric as calculated in \cite{alvi,alvi1}. This method makes use of a mathematical technique called asymptotic matching. Alvi \cite{alvi,alvi1} was the first person to attempt such a construction. The four-dimensional global metric is subdivided into four regions, each with their own approximation schemes for solving the Einstein equations, which are then \textit{`glued'} together in a \textit{buffer zone} in which two adjacent metric approximations are valid. The work by Alvi is based on earlier research by Manasse \cite{manasse} and D'Eath \cite{d'eath,d'eath1}. 

The main goal of the broader project we consider is to determine whether a mechanism exists that will describe the fate of particle trajectories existing in a spacetime around merging binary black hole systems. We want to identify potential mergers before inspiral. Therefore, we want to find a characteristic signal unique in its identity that could alert one to the stage at which the particles are ejected, engulfed or has become faint and inactive. This would consequently give us an early warning of gravitational wave emission from the inspiral. In this proceeding we tackle only part of the problem. We ignore, leaving to future research, the angular momentum accompanying spinning black holes. 

Making use of the analytical spacetime from \cite{alvi,alvi1}, we propose to study the geodesics living on this spacetime. Geodesics are simple to study and they allow us to track the movement of particles in the spacetime. This will allow us to determine how many particles are falling into the inner zone of the binary and eventually the black holes.

The structure of this paper is as follows: Section 2 concerns the structure of the spacetime we use and its limitations. Section 3 deals with the problem of computing and using geodesics in a numerical simulation. Then, Section 4 deals with the discussion of our results. Finally, we end with the conclusion.


\section{Spacetime structure}
\begin{center}
	\begin{figure}[ht!]
		\centering
		\resizebox{0.7\hsize}{!}{\includegraphics{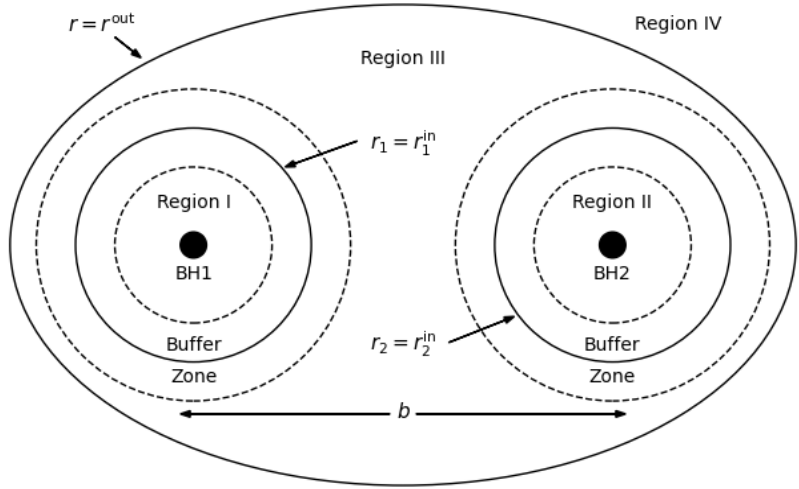}}
		\caption{Global metric displaying all regions following (gr-qc/991211).}
		\label{full}
	\end{figure}
\end{center}

Our project begins by considering two equal mass ($m_1 = m_2 = 30 M_{\odot}$) Schwarzschild black holes, each of them perturbed by the other, i.e. each black hole exerts a tidal force on the other. They are widely separated ($\sim 14 M$ in geometric units where $M = m_1 + m_2$), this allows for the subdivision of the spacetime. For $14 M$ separation, the black hole speeds are $\frac{v}{c} \approx 0.13$, thus being on the very edge of the slow-motion regime. The analytic metric of \cite{alvi,alvi1} generates a space-like hypersurface from the full spacetime. The global metric on each hypersurface is made up of 4 individual regions (see Figure~\ref{full}), all held together by asymptotically \textit{patching} in the buffer zones, which cover the space where two adjacent regional metrics overlap and are simultaneously valid. In patching, one sets the metrics in the buffer zone equal to each other at a point, i.e. set approximate solutions to the Einstein equations equal to each other on specified 2-surfaces \cite{yunes}. 

In regions I and II, the inner zones, the perturbed black hole metrics are patched to region III, the near zone. Region III is described by a post-Newtonian approximation which holds for the weak-field ($GM/(rc^2) << 1$) and slow-motion ($v/c << 1$) limit\footnote{Here \textit{c} and \textit{G} are the speed of light and Newton's gravitational constant, whilst \textit{v, M, r} are the characteristic velocity, mass and separation of the system \cite{yunes}.} \cite{blanchet}. A post-Minkowski metric describes region IV, the far zone. Alvi considers the near and far zones as already patched by construction.

In order to study geodesics, we need to produce a full spacetime from these hypersurface slices. The slicing that allows for the production of the hypersurfaces remains valid when the separation is $> 10 M$ (along with the regional decomposition) \cite{Mundim_2014}. Thus, our approach is to generate a full spacetime out of a collection of space-like hypersurfaces each corresponding to a given time. The evolution of the binary itself provides the evolutionary linkage between each hypersurface. This is because the analytic metric depends on the separation and speed of the binary, on a given hypersurface, and we impose these evolutions via the relations from \cite{Isoyama:2020lls} according to the time assigned to that hypersurface. 

\section{Calculating geodesics}
Our results are based on the calculation of geodesics. This is a very useful probe, since it provides a relatively simple way to study the motion of particles on different trajectories as well as their fate. As mentioned already, the metric we used to get our geodesics was an analytic metric by Alvi. To obtain geodesics, we needed to compute the Christoffel symbols in this spacetime first. We compute these symbols using finite differencing techniques up to second order. Our geodesics were computed with the time coordinate as a parameter since this formalism is useful in numerical computations:
\begin{equation}\label{eq}
\frac{d^2 x^{\mu}}{d t^2} = -\Gamma^{\mu}_{\alpha\beta} \frac{d x^{\alpha}}{dt}\frac{d x^{\beta}}{dt} + \Gamma^{0}_{\alpha\beta} \frac{d x^{\alpha}}{dt}\frac{d x^{\beta}}{dt}\frac{d x^{\mu}}{dt}.
\end{equation} 
Furthermore, we did this using the Runge–Kutta–Fehlberg~\cite{fehlberg} method. Then, using the geodesic equation, we can produce acceleration maps which provide insight into where electromagnetic radiation would be most intense. However, the actual geodesics themselves belong to the particles in the spacetime. Thus, we can observe the forces acting on particles in the disk without employing fluid dynamic methods.

\section{Results and discussion}   
In this work we have computed the trajectories of particles in a four-dimensional, subdivided spacetime made up of four regions held together by asymptotic patching in the buffer zones containing adjacent metrics. This means that adjacent metrics are set equal to each other at a point, rather than in the entire buffer zone region \cite{yunes}. This means that the global metric is mildly discontinuous and errors may be introduced. 

These analytic metrics were designed to calculate initial data for binary black hole simulations. Hence, we note that Alvi mentions that before one extracts initial data using this metric, the discontinuities need to be smoothed out. Yunes et. al. \cite{yunes1} found a way to deal with the issues in Alvi by creating transition functions \cite{yunes2} which broke the discontinuities and allowed transitions from one region to its adjacent region without the introduction of errors.

\begin{center}
	\begin{figure}[ht!]
		\centering
		\includegraphics[scale=0.6]{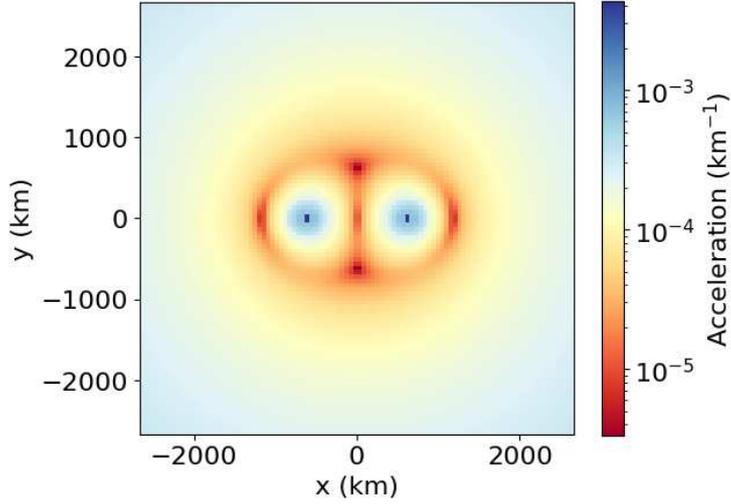}
		\caption{Acceleration map of the whole global metric.}
		\label{fullAcc}
	\end{figure}
\end{center}

Let us begin by addressing Figure~\ref{full}. This shows the full global metric in Alvi \cite{alvi} and allows one to visualize how the regions are related. One thing we need to be cautious about is that Figure~\ref{full} is somewhat misleading. In that figure all regions seem to be connected and even appear to have smooth transitions between each of the zones. However, the full metric contains discontinuities at the boundaries between regions. Despite this, Figure~\ref{fullAcc} shows no visible signs of discontinuity in the accelerations. 

Next, we consider each region of the global metric separately. The acceleration maps produced allow one to visualise the areas of the region where electromagnetic radiation is likely to be most intense.

\subsection{Regions I and II}
\begin{center}
	\begin{figure}[ht!]
		\centering
		\resizebox{0.49\hsize}{!}{\includegraphics{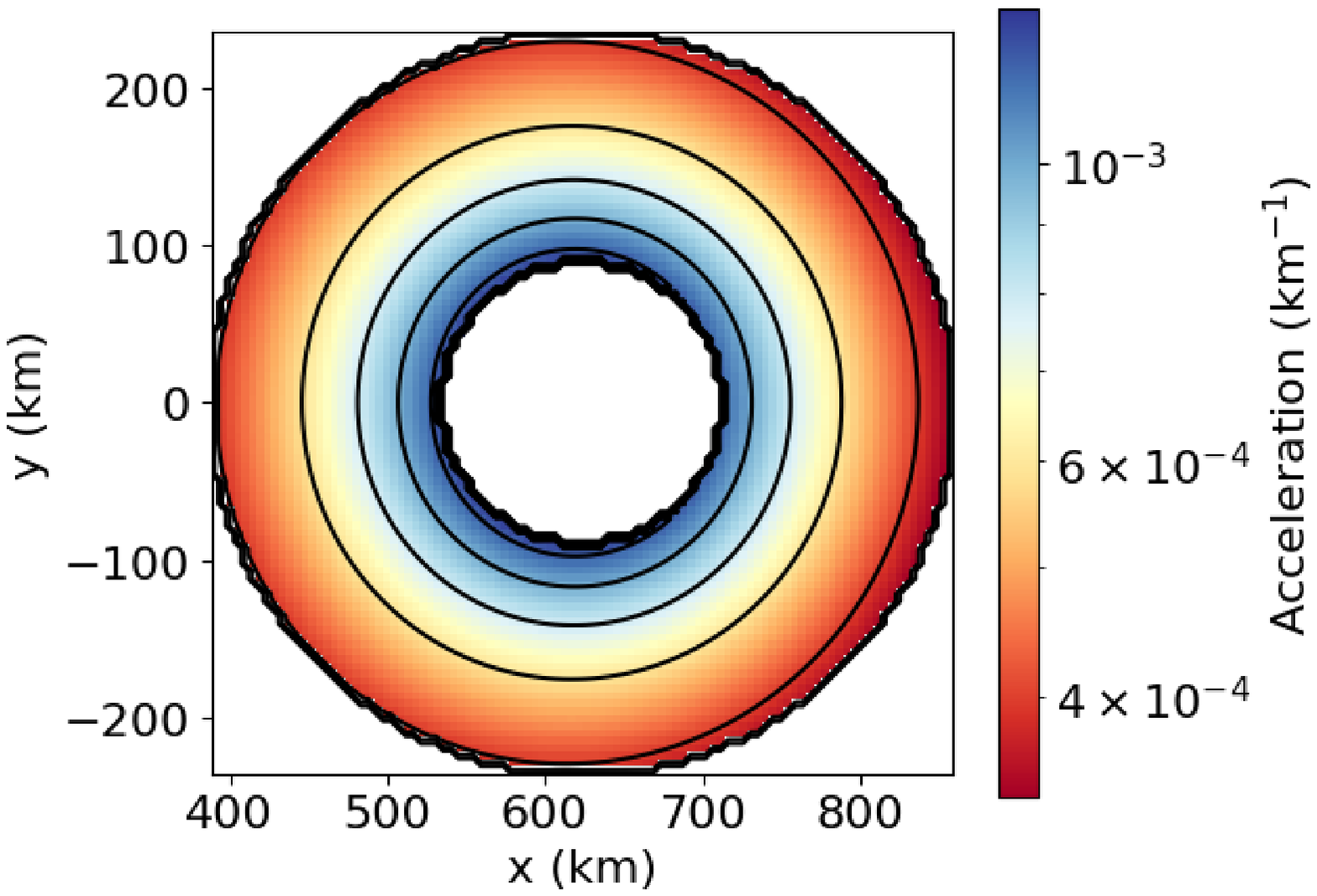}}
		\resizebox{0.49\hsize}{!}{\includegraphics{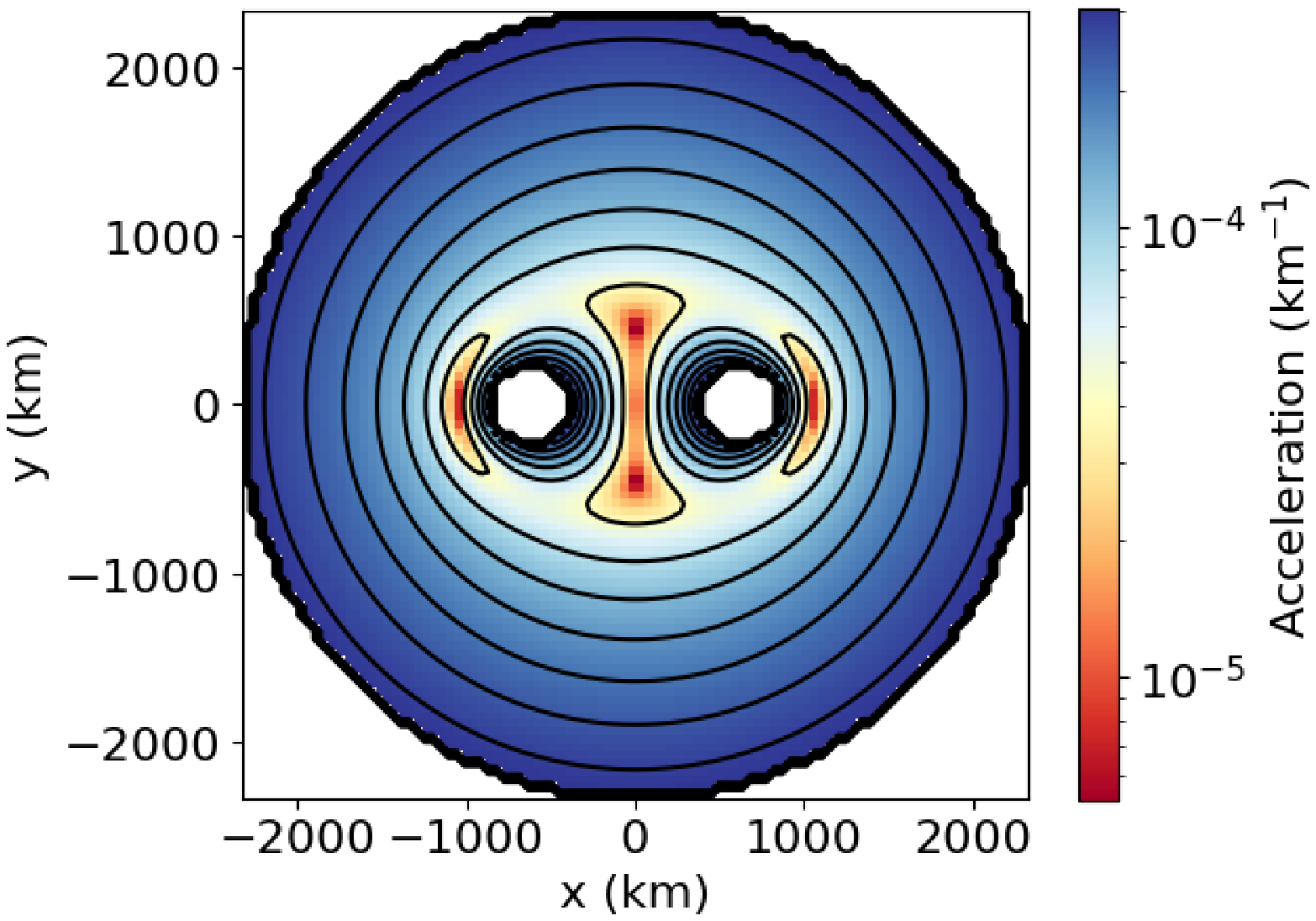}}
		\caption{Left: the acceleration map covering Regions I and II which surround BH 1 and 2, respectively. Right: acceleration map covering Region III.}
		\label{inner}
	\end{figure}
\end{center}

The left panel of Figure~\ref{inner} shows the inner regions around black hole 1 and black hole 2. The black holes are of equal mass. We have excised regions that we are not discussing (masked in white). For the case we now consider, we see that the acceleration is larger nearer to the black holes. This is expected since the gravitational pull is larger there. Additionally, the contour lines circularize as they move away from the black holes due to the perturbation.

\subsection{Region III}
This region, also called the near zone, shows how particles accelerate around both black holes when dealing with a centre of mass problem. Here, in the right panel of Figure~\ref{inner}, we have excised the inner regions and the far zone. The centre shows several nodal regions where the acceleration is very small relative to the rest of the space. We point out that the acceleration gets larger as we move away from the centre, excised regions. This is due to the use of co-rotating coordinates. Note that we validated our algorithm on a Schwarzschild black hole metric to ensure this effect was not spurious.

\subsection{Geodesics}
We display geodesics for 100 initially stationary particles starting in region III, they have a maximum simulation time of $\approx 0.1$ s. This duration is such that the binary remains widely separated throughout, the time where the Schwarzschild radii would cross being $\approx 0.89$ s. To justify our choice of stationary initial conditions, we note the particles of interest are coming from a circumbinary, accretion disk. Therefore, the maximum initial speed of these particles is the speed of sound in the disk: $v_s = \sqrt{\frac{3}{8}} \omega z_0$~\cite{pringle}, where $z_0 \approx 1 M$~\cite{Gold} is half the disk thickness and we have assumed a polytropic disk with index $4/3$~\cite{Gold}. In the co-rotating coordinates, the initial speed of a stationary particle is $\sim \omega r$. Comparing this to $v_s$ we find $\frac{v_s}{\omega r} \approx 0.035$ when $r \sim 1000$ km. Thus, an initial orbital velocity is a small correction which we ignore as a first order approximation.
In our plots geodesic origin points are given by `star' markers. We will use a black ellipse surrounding the geodesics to give the border of the near zone. 

\begin{center}
	\begin{figure}[ht!]
		\centering
		\resizebox{0.49\hsize}{!}{\includegraphics{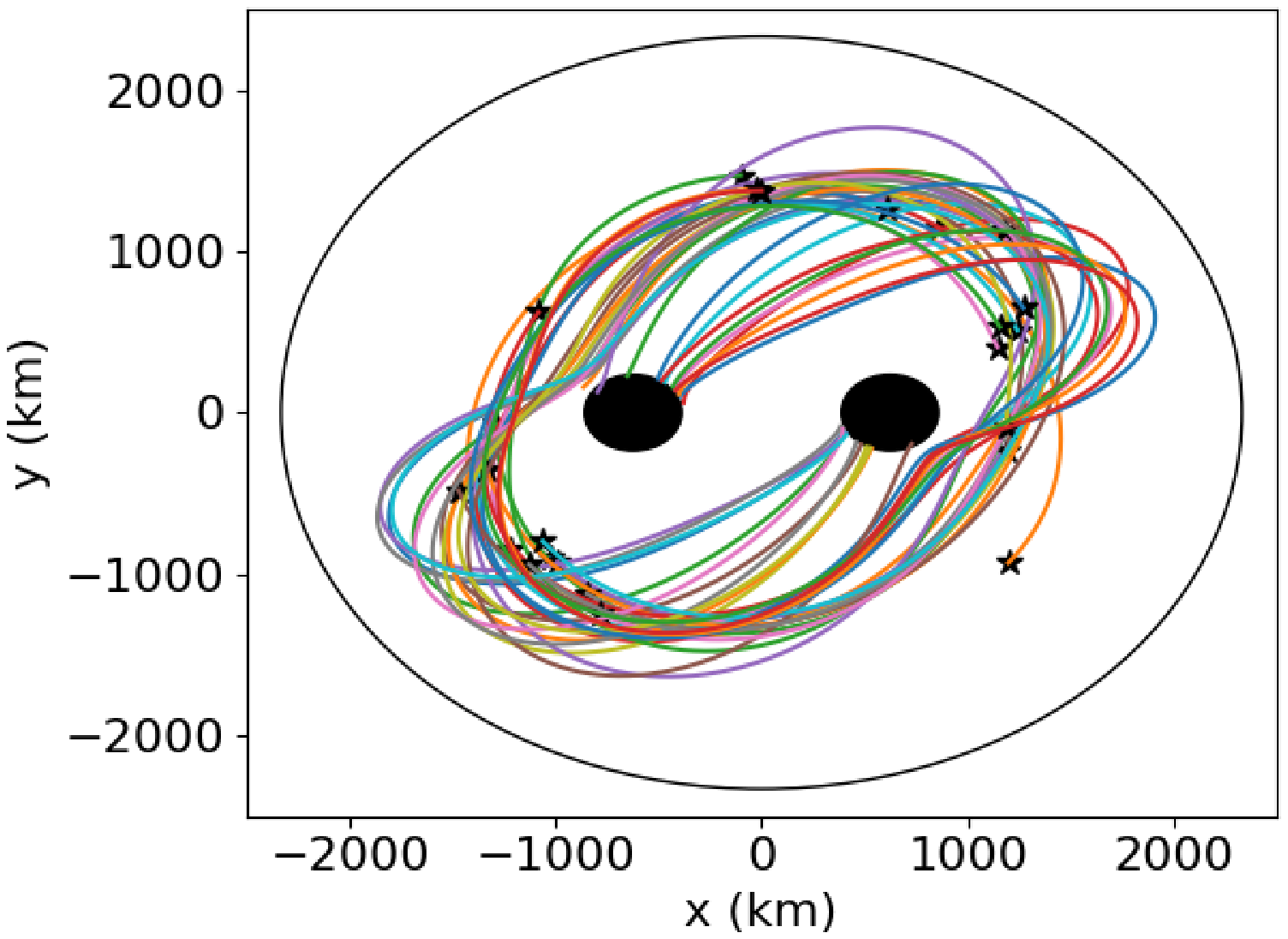}}
		\resizebox{0.49\hsize}{!}{\includegraphics{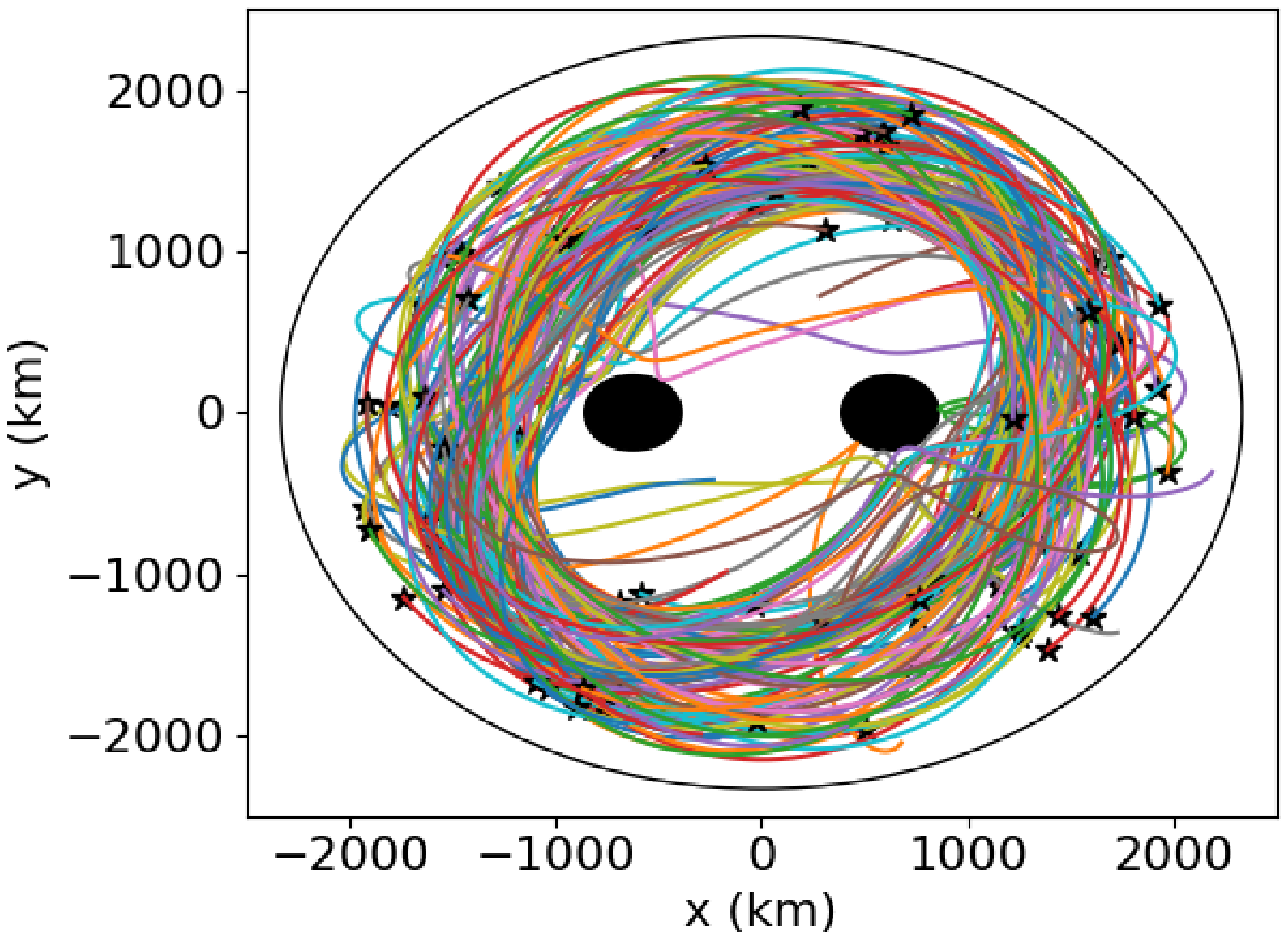}}
		\caption{Geodesics for potentially accreting geodesics (left) and those that do not accrete (right).}
		\label{geo}
	\end{figure}
\end{center}

Figure~\ref{geo} shows both the geodesics that will potentially accrete onto the black holes and those that do not within the  maximum simulation time of $\approx 0.1$ s. It is notable that around $24$\% of the particles fall into the black holes within the simulation time.

\section{Conclusions}
Studying geodesics, computationally, provides a unique means to search for precursor electromagnetic signals that would herald the release of gravitational waves prior to the tight inspiral phase of the binary. If the environment around the black holes is magnetised, synchrotron radiation is likely to occur with any flow of particles towards the black holes, producing radio emissions which may be detectable by MeerKAT. Though this is more probable for larger black holes than studied here as such small regions may not be resolvable. Magnetized environments will also require additional dynamical corrections, as strong magnetic fields will affect the particle geodesics. The effect of such fields will be considered in future work.

We have assumed that a disk, straddling the near zone \cite{shapiro1}, exists. It is contained within a spacetime described by an analytic metric. We observed the motion of individual particles by computing the geodesics of each particle in the spacetime and found initial indicators that if the particles begin from rest in the near zone, they can be rapidly swallowed up by the black holes. In addition, the orbits within the near zone appear to be unstable. The instability observed does not occur due to effects resulting from the individual black holes as the particles starting closest to an inner zone can be found at $r\gtrsim 3 r_{\rm ISCO}$ from that black hole. 

\section{Acknowledgements}
We thank the referee, whose comments greatly improved the presentation of this work. JT acknowledges the support of the SARAO post-doctoral fellowship initiative. GB acknowledges the financial support of the National Research Foundation of South Africa via Thuthuka grant no. 117969.

\section*{References}

\medskip

\bibliographystyle{iopart-num}
\bibliography{bib}

\providecommand{\newblock}{}
\begin{thebibliography}{10}
\expandafter\ifx\csname url\endcsname\relax
  \def\url#1{{\tt #1}}\fi
\expandafter\ifx\csname urlprefix\endcsname\relax\def\urlprefix{URL }\fi
\providecommand{\eprint}[2][]{\url{#2}}

\bibitem{Mundim_2014}
Mundim B~C, Nakano H, Yunes N, Campanelli M, Noble S~C and Zlochower Y 2014
  {\em Phys. Rev. D\/} {\bf 89} 084008

\bibitem{baumgarte}
Baumgarte T and Shapiro S 2010

\bibitem{cook}
Cook G~B 2000 {\em Living Rev. Rel.\/} {\bf 3} 5

\bibitem{alvi}
Alvi K 2000 {\em Phys. Rev. D\/} {\bf 61}(12) 124013

\bibitem{alvi1}
Alvi K 2003 {\em Phys. Rev. D\/} {\bf 67}(10) 104006

\bibitem{manasse}
Manasse F~K 1963 {\em J. Math. Phys.\/} {\bf 4} 746--761

\bibitem{d'eath}
D'Eath P~D 1975 {\em Phys. Rev. D\/} {\bf 11}(6) 1387--1403

\bibitem{d'eath1}
D'Eath P~D 1975 {\em Phys. Rev. D\/} {\bf 12}(8) 2183--2199

\bibitem{yunes}
Yunes N, Tichy W, Owen B~J and Br\"ugmann B 2006 {\em Phys. Rev. D\/} {\bf
  74}(10) 104011

\bibitem{blanchet}
Blanchet L 2014 {\em Living Rev. Rel.\/} {\bf 17} 2

\bibitem{Isoyama:2020lls}
Isoyama S, Sturani R and Nakano H 2020

\bibitem{fehlberg}
Fehlberg E 1969 {\em NASA\/} {\bf 315}

\bibitem{yunes1}
Yunes N and Tichy W 2006 {\em Phys. Rev. D\/} {\bf 74} 064013

\bibitem{yunes2}
Yunes N 2007 {\em Clas. and Quant. Grav.\/} {\bf 24} 4313–4336

\bibitem{pringle}
Pringle J~E 1981 {\em Annu. Rev. Astron. Astrophys\/} {\bf 19} 137--160

\bibitem{Gold}
Gold R, Paschalidis V, Etienne Z~B, Shapiro S~L and Pfeiffer H~P 2014 {\em
  Phys. Rev. D\/} {\bf 89} 064060

\bibitem{shapiro1}
Khan A, Paschalidis V, Ruiz M and Shapiro S~L 2018 {\em Phys. Rev. D\/} {\bf
  97}(4) 044036

\end{thebibliography}

\end{document}